\documentclass[preprint]{aastex}

\def \cm{~\rm{cm}}
\def \s{~\rm{s}}
\def \km{~\rm{km}}

\def \K{~\rm{K}}
\def \g{~\rm{g}}

\def \AU{~\rm{AU}}

\def \yr{~\rm{yr}}

\def \astrobj#1{#1}
\begin{document}
%\small

\title{THE ROLE OF GRAVITY IN WINDS COLLISION IN THE ETA CARINAE BINARY SYSTEM}

\author{Muhammad Akashi\altaffilmark{1} and Noam Soker\altaffilmark{1}}
\altaffiltext{1}{Department of Physics, Technion$-$Israel Institute of Technology,
  Haifa 32000 Israel; akashi@physics.technion.ac.il;
  soker@physics.technion.ac.il.}

\begin{abstract}
We conduct 3D numerical simulations of the winds collision process in the massive binary
system $\eta$ Carinae, and conclude that accretion occurs during periastron passage.
We include radiative cooling of the two winds, one from each star, and the gravity of
the secondary star.
Our new numerical finding is that at an orbital separation of $r \sim 3-4 \AU$,
about three weeks before periastron passage, accretion of dense primary wind gas onto
the secondary star begins.
To isolate the basic role of the secondary stellar gravity, we neglect the orbital
motion and the acceleration zone of the primary wind.
Including these effects will strengthen even more our conclusion that accretion
near periastron passage of $\eta$ Car is inevitable.
Accretion of the primary wind gas onto the secondary star for several weeks near
periastron passage accounts for the otherwise puzzling behavior of the binary system
near periastron passage.
\end{abstract}

%  \keywords{ (stars:) binaries: general$-$stars: mass loss$-$stars:
%  winds, outflows$-$stars: individual ($\eta$ Car) }

%\twocolumn

% ==========================================================
\section{INTRODUCTION}
\label{sec:intro}
% ==========================================================

\astrobj{$\eta$ Car} is a very massive binary system
with a well established orbital period of $5.54 \yr$
(Damineli et al. 2008a; Fernandez Lajus et al. 2009;
Landes \& Fitzgerald 2009). However, many other parameters
of the orbital motion and the behavior of the stars and
their winds are not well determined, and sometimes are in dispute.
Examples of not well determined parameters are the masses of the two stars,
with values that can be found in the literature in the ranges
$100 \la M_1 \la  200 M_\odot$ and $20 \la M_2 \la  80 M_\odot$ for the primary
and secondary stars, respectively.
The lower limits come from the Eddington luminosity
limit, and the upper values from analysis of the Great
Eruption of the 19th century and fitting to stellar
evolution tracks (Kashi \& Soker 2010).
An example for a parameter in a fierce dispute
is the orientation of the semi major axis of the orbit,
with practically opposite values in the literature
(e.g., Nielsen et al. 2007; Damineli et al. 2008b; Falceta-Goncalves et al. 2005;
Abraham et al. 2005; Kashi \& Soker 2007b, 2008a, 2009b,c; Davidson 1997; Smith et al. 2004;
Henley et al. 2008; Abraham \& Falceta-Goncalves 2007, 2010;
Gull et al. 2009; Groh et al. 2010; Richardson et al. 2010; Mehner et al. 2010).

Some major puzzles are connected to the process that
causes the 4-10 weeks long decrease in the X-ray
emission during each periastron passage (Corcoran 2005; Hamaguchi et al. 2007).
The long duration of the deep X-ray minimum cannot be
explained by an eclipse or absorption, and it imposes strong
constrains on the properties of the binary system (Kashi \& Soker 2009a,c).
The source of the hard X-ray emission is the shocked
wind of the secondary star (Corcoran et al. 2001; Pittard \&
Corcoran 2002; Akashi et al. 2006). The only successful
explanation for the long X-ray minimum is the accretion model.
In this model for several weeks the less massive
secondary star is accreting mass from the primary wind
instead of blowing its fast wind. The absence of the
secondary wind explains the X-ray minimum.
Despite that the accretion model was successfully
developed in a series of papers (Soker 2005; Kashi \& Soker
2008b, 2009a, to list a few of them), not all researchers
are convinced that accretion takes place indeed.
Numerical simulations of the two winds collision in $\eta$ Car did
not include the gravity of the secondary star,
and could not obtain an accretion (Pittard et al. 1998;
Pittard \& Corcoran 2002; Okazaki et al. 2008; Parkin et al. 2009).
Numerical simulations that did include the gravity of the stars
(Pittard 1998; Stevens \& Pollock 1994) had parameters that are not
relevant to $\eta$ Car, and did not consider accretion.

In the present study we conduct numerical simulations
of the collision of the two winds in
the $\eta$ Car binary system. We perform Cartesian 3D
numerical simulations.
Our results suggest that accretion is very likely to occur near periastron passage.

% ========================================
\section{NUMERICAL SET UP AND ASSUMPTIONS}
\label{sec:numeric}
% ========================================

% -----------------
\subsection{The numerical method and parameters}
\label{sec:parameters}
% -----------------
There are some uncertainties as to the exact binary
parameters of $\eta$ Car (Kashi \& Soker 2009c). We
here take the common parameters as used by many others,
e.g., list of papers in section 1.
The  stellar masses are $M1 = 120 M_\odot$, and $M2 = 30 M_\odot$,
the eccentricity is $e = 0.9$, and the orbital period is $2024$ days,
hence the semi-major axis is $a = 16.64 \AU$ and periastron occurs at $r = 1.66 \AU$.
The mass loss rates and velocities of the winds are $\dot M_1= 3 \times 10^{-4} M_\odot \yr^{-1}$ and
$\dot M_2=  10^{-5} M_\odot \yr^{-1}$, and $v_1 = 500 \km \s^{-1}$
and $v_2 = 3000 \km \s^{-1}$, respectively.

The simulations are performed with Virginia Hydrodynamics-I
(VH-1), a high resolution multidimensional astrophysical
hydrodynamics code developed by John Blondin and co-workers
(Blondin et al. 1990; Stevens et al., 1992; Blondin 1994). We have
added radiative cooling to the code at all temperatures of $T > 2
\times 10^4 \K$. Namely, the radiative cooling is set to zero for
temperatures of $T<2 \times 10^4 \K$, but we allow adiabatic
cooling to continue below that temperature. Radiative cooling is
carefully treated near contact discontinuities, to prevent large
temperature gradients from causing unphysical results. The cooling
function $\Lambda (T)$ (for solar abundances) is taken from
Sutherland \& Dopita (1993; their table 6). Gravity by the
secondary star is included, as this is the sole issue of this study.
In two runs we added the gravity of the primary star, and show it cannot prevent accretion.
However, as the acceleration zone of the primary star is large and we cannot add radiative pressure,
the primary gravity is not fully self-consistent.

We start by imposing undisturbed winds in the numerical
grid, and let the flow reach a steady state without gravity. Only
then we turn on gravity, and let the flow reach the new steady
state with the gravity included. The flow does not reach a
strict steady state, but rather it posses an erratic motion of the
winds collision region that we term wiggling. The collision region
moves back and forth, i.e., its distance from the stars is not
constant, and to the sides, i.e., the axi-symmetry around the line
connecting the two stars is broken.

We perform the numerical simulations in the Cartesian geometry $(x,y,z)$ mode
of the code (a 3D calculation).
Results are presented in the (x,y) plane that contains the two stars.
In most runs there were 112 equal-size grid points along each axis.
The distance between the two stars is half the length of the x-axis.
To confirm the adequacy of the resolution, we run one case with gravity
and an orbital separation of $r=2 \AU$ and with 175 cells along each axis (instead of 112).
We found the differences from the lower resolution run to be small.

In the present study we aim at understanding the basic role of gravity, neglecting the role
of orbital motion and the acceleration zone of the primary wind.
Including these two effects will increase the accretion rate and makes it
start earlier even (see section \ref{sec:assumptions}).
We set the distance between the stars to be constant at each numerical run,
and let the flow reach a, more or less, steady state.
It is not a strict steady state, as the winds
interaction zone is wiggling, as was already found by Pittard et al. (1998),
Pittard \& Corcoran (2002), Okazaki et al. (2008), and Parkin et al. (2009).
We conduct the runs with and without the secondary
stellar gravity, and for several orbital separations.

At the location of each star we inject its appropriate wind.
The winds were injected from a square of size $8\times 8$
cells around each star.
Namely, 4 cells from the star along the axes.
This prevents the accreting primary wind to
reach a distance closer than 5 cells from the
center of the secondary (as at each time step the
condition of outflow is imposed there).
As in most runs there are 56 cells between the
stars, this closest distance is $\sim 0.1 r$.
As the condition of accretion we take the presence
of dense primary wind gas at a distance of 5 cells ($0.1 r$) from the center of the secondary star.

The boundary conditions of the simulation box are outflow at the $6$ sides of the box;
the gas that flows out of the box cannot flow back into the box.

% -----------------
\subsection{Assumptions and approximations}
\label{sec:assumptions}
% -----------------
Our calculations have the aim of revealing the role of gravity in the winds
collision process when $\eta$ Car approaches its
periastron; near apastron the role of gravity is small.
For that we varied only the distance between the
stars (orbital separation), and let the flow reach a steady state (up to a wiggling motion)
with and without gravity.
We now discuss the implications of the main
assumptions and approximations.

(1) \emph{Acceleration zone of the primary wind.}
While we inject the primary
wind at its terminal velocity of $v_1=500 \km \s^{-1}$,
the winds from OB super-giants are accelerated over a large distance of
about several stellar radii.
A commonly fitting formula at several stellar radii is
$v \simeq v_1[1-(R_1/r)]^{\beta}$, with $\beta \simeq 1-3.5$
(e.g., Kraus et al. 2007, and references therein).
If, for example, we take $R_1=0.8 \AU$ and $\beta=2$, we find that the
wind velocity is $v(2 {\AU})=0.36 v_1$ (just
before periastron) and $v(4 {\AU})=0.64 v_1$.
The density is $v_1/v$ times as high compared
with the value according to our
assumption of a constant wind speed.
When the orbital motion is considered, it is
found that the stagnation point
of the colliding winds moves toward the
secondary, and by that strengthening the accretion process.
Also, when the acceleration zone is considered, the accretion rate is increased
by a substantial fraction (Kashi \& Soker 2009c).
Neglecting the acceleration zone is our strongest assumption.
Including it will make our conclusions that accretion is inevitable stronger even.

(2) \emph{Clumping in the primary wind.}
The primary wind is most likely clumpy (Moffat \& Corcoran 2009).
We inject a smooth primary wind, and neglect clumping.
Clumps can more easily penetrate the secondary wind (Soker 2005), and might start to
be accreted earlier even than what our results indicate.

(3) \emph{Orbital motion.} (3.1) The first
effect is that there is no cylindrical symmetry in the flow any more.
The effect is most pronounced at large distances from the
stagnation point (Okazaki et al. 2008).
Since accretion starts from regions near the stagnation
point, deviation from axi-symmetry is not a major issue.
\newline
(3.2) The inclusion of orbital motion increases
the ram pressure of the primary wind before periastron, and will make our
results more robust even.
For example, about a month before periastron passage the orbital separation is $r=4 \AU$,
and the two stars relative velocity is $v_o \simeq 200 \km \s^{-1}$.
Inclusion of this velocity will support the accretion process.
The positive role of the orbital motion is most pronounced when the acceleration zone
of the primary wind is also considered.
\newline
(3.3) The orbital motion implies that our assumption of a steady state brakes down.
We find that when the orbital separation is $r \simeq 3 \AU$ accretion starts.
Including the effect discussed above accretion will start at $r \sim 4 \AU$.
At $r=4 \AU$ the orbital time scale is
$t_o \equiv r/v_o \sim 1~$month.
The accretion time is the time the primary wind crosses the distance from
the stagnation point to the secondary. We find this time
to be $\sim 0.25r/v_1 \simeq 0.1 t_o$.
Therefore, the effect of deviation from a steady state when we find accretion to
start is very small.

(4) \emph{Secondary Radiation pressure.}
For the parameters used here the radiation pressure is $0.6$ times the ram
pressure of the secondary wind (Soker 2005).
Therefore, the effect of radiation pressure
 would be like increasing the ram pressure
of the secondary wind by a factor of 1.6.
But as we neglect the orbital motion, that
would increase the ram pressure of the primary
wind by a similar factor when accretion starts,
 we can safely neglect the secondary radiation pressure.
That the secondary radiation cannot prevent
accretion was discussed in length by Soker (2005).
In any case, we performed one simulation with the secondary mass loss rate
set at 1.6 times the regular value.
We still obtained accretion when the orbital separation is $r=3 \AU$.

(5) \emph{Magnetic fields in the primary wind.} The post shock
primary wind radiatively cools very fast. It is therefore
compressed to high densities by the surrounding pressure. The
compression might amplify an initially weak magnetic field to
become dynamically important (Kashi \& Soker 2007a). As discussed
by Kashi \& Soker (2007a), the magnetic fields in the postshock
primary wind limit the compression of the postshock primary wind.
This influences the flow structure and the recombination rate of
the postshock primary wind. We neglect the effect of magnetic
fields.

% ========================================
\section{RESULTS}
\label{sec:results}
% ========================================

We will present the results as temperature, density, and velocity maps
for the different cases.
It should be noted that for each case the maps are given at a particular time.
Although the flow reaches a general steady state, it wiggles around that
average steady state (like a flag in a wind; Pittard et al. 1998;
Pittard \& Corcoran 2002; Okazaki et al. 2008; Parkin et al. 2009).
The erratic behavior of the flow appears also in the deviation from initial axisymmetry condition,
as any tiny numerical noise is amplified to a large departure from symmetry.
In the figures this will be noticed as an asymmetry of the upper and lower parts.
We will demonstrate this for the case of an orbital separation of $r=2 \AU$.
We also recall that the accretion condition is that the dense primary gas reaches the zone
of injection of the secondary wind. For most runs this is at
$4+1~{\rm cells}~=0.1r$.
In Table 1 we summarize the cases we run.
% TTTTTTTTTTTTTTTTTTTTTTTTTTTTTTTTTTTTTTTTTTTTTTTTT
\begin{table}

Table 1: Simulated cases

\bigskip
\begin{tabular}{|l|c|c|c|c|c|}
\hline
 Run & $r (\AU)$ & $M_2$ Gravity  & Accretion & comments & Figure \\
\hline
2G   &  2 & Yes & Yes &  &  \ref{fig:fig2yes}+\ref{fig:fig2yest}+\ref{fig:temps_map}a\\    % Fig 1 + 2 + 9
\hline
2GH  & 2  & Yes & Yes & High resolution & \ref{fig:fig2yesH} \\        %  Fig  3
\hline
2N   &  2 & No  & No &  & \ref{fig:fig2no}\\                           %  Fig 4
\hline
3G   &  3 & Yes & Yes & & \ref{fig:fig3yes}\\                          %  Fig 5
\hline
4G   &  4 & Yes & No &  & \ref{fig:fig4yes} +\ref{fig:temps_map}b \\                         %  Fig 6
\hline
10G  & 10 & Yes & No & & \ref{fig:fig10yesno}a +\ref{fig:temps_map}c  \\                 %  Fig 7
\hline
10N  & 10 & No  & No & & \ref{fig:fig10yesno}b \\                 %  Fig 7
\hline
3GR  &  3 & Yes & Yes & $1.6 \dot M_2$ & \ref{fig:fig2yes16}\\          %  Fig 8
\hline
2GG  &  2 & Yes & Yes & $M_1$ Gravity  &  \ref{fig:gravity1}a \\    % Fig 10a
\hline
3GG  &  3 & Yes & Yes & $M_1$ Gravity  &  \ref{fig:gravity1}b \\    % Fig 10b
\hline
\end{tabular}

\footnotesize
\bigskip
For all runs but one, the mass loss rates and velocities of the winds are
$\dot M_1= 3 \times 10^{-4} M_\odot \yr^{-1}$ and
$\dot M_2=  10^{-5} M_\odot \yr^{-1}$, and $v_1 = 500 \km \s^{-1}$
and $v_2 = 3000 \km \s^{-1}$, respectively.
In run 3GR the mass loss rate from the secondary is 1.6 as high to mimic
the secondary radiation pressure.
In all runs but one, the number of cells along each axis is 112, with 56 cells between the two
stars, and the winds are injected to the grid at 4 cells from each star.
In run 2GH these numbers are 175, 88, and 4,  respectively.
\normalsize
\end{table}
% TTTTTTTTTTTTTTTTTTTTTTTTTTTTTTTTTTTTTTTTTTTTTTTTT

\begin{figure}
    \includegraphics[scale=0.6]{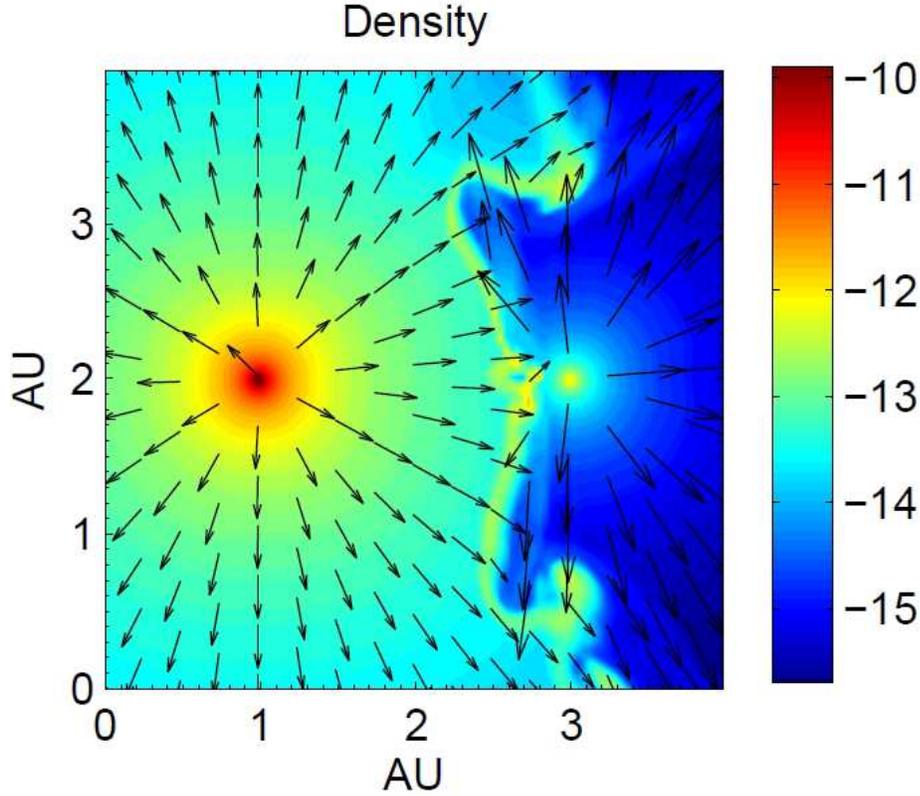}
        \caption{Density and velocity maps for an orbital separation of $r=2 \AU$, and
        the gravity of the secondary star included.
        The bar on the right gives the density color-code in units of $\log[\rho(\g \cm^{-1})]$.
        The arrows are scaled in four intervals, from the shorter to the
        longer arrow, according to:
        $0 < v < 400 \km \s^{-1}$,$400 < v < 800 \km \s^{-1}$,
        $800 < v < 1600 \km \s^{-1}$, and $v > 1600 \km \s^{-1}$ .
        The secondary is the star on the right hand side.
        A dense blob is clearly seen at a distance of
        $0.1 r$ from the secondary. This is our condition
        for accretion, as this is the distance where
        the secondary wind is injected. }
   \label{fig:fig2yes}
     \end{figure}

\begin{figure}
   \includegraphics[scale=0.6]{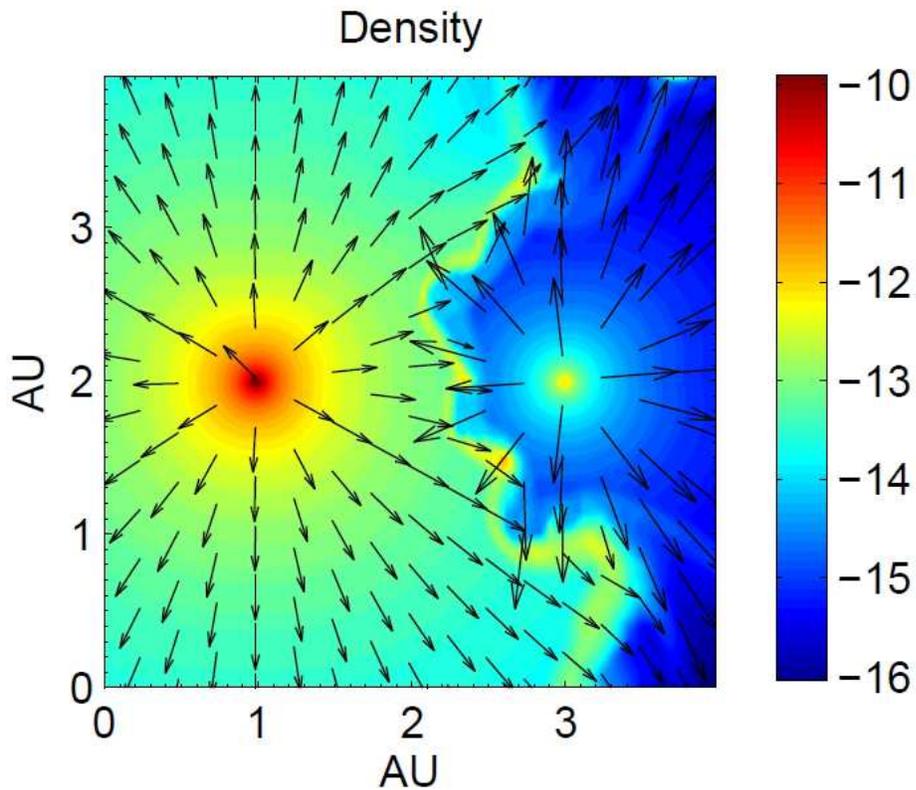}
        \caption{Like Fig. \ref{fig:fig2yes} but
        at nearly one month later.
        This demonstrates the wiggling behavior of
        the winds collision process, and our imposed conditions near the secondary star.
        If instead of imposing outflowing wind near the secondary, we would let the
        primary wind to reach the secondary itself, we would had a
        continuous accretion.
        Our limited resolution does not allow us to do so as we do not resolve the acceleration
        zone of the secondary wind. }
   \label{fig:fig2yest}
     \end{figure}

\begin{figure}
   \includegraphics[scale=0.6]{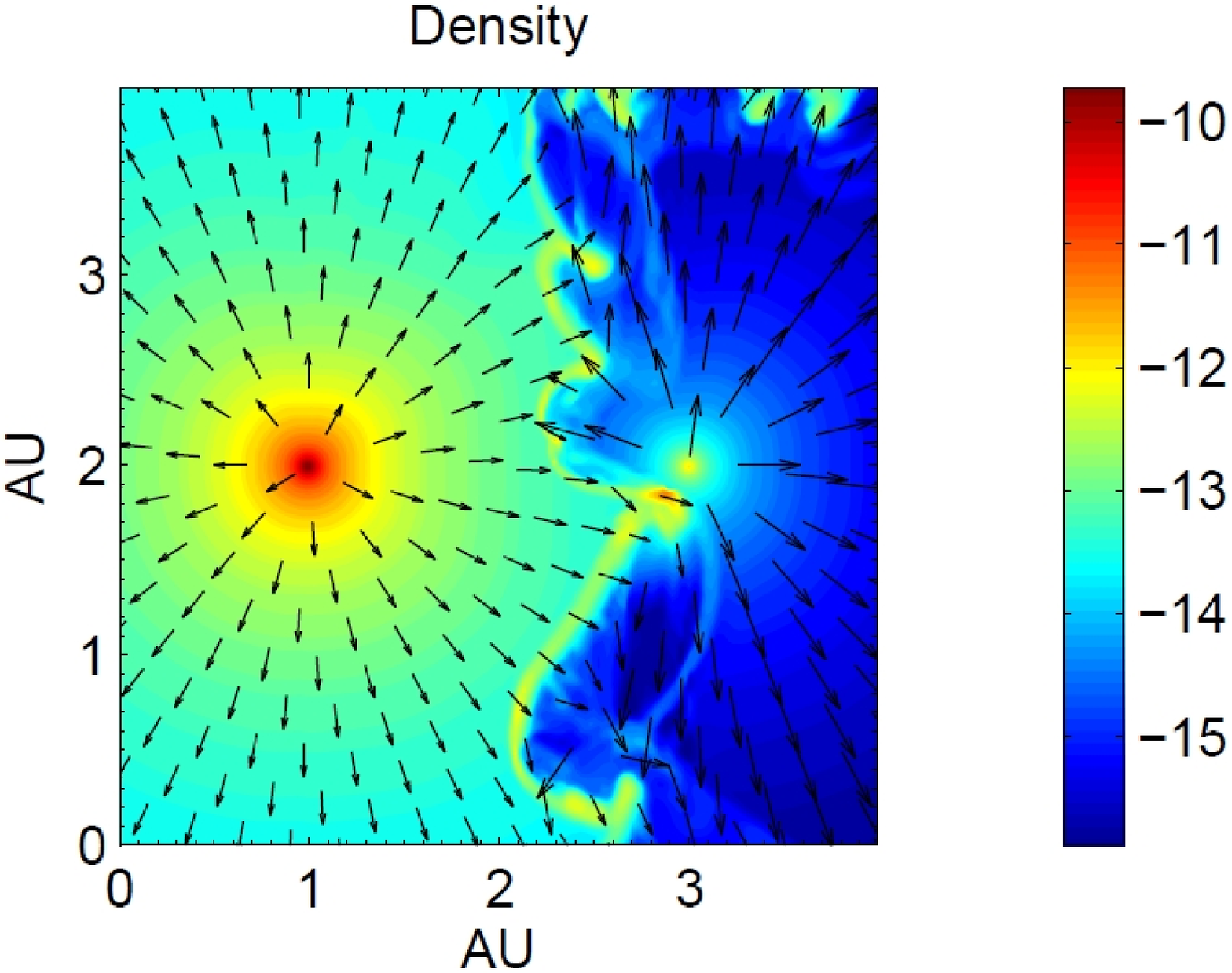}
        \caption{Like Fig. \ref{fig:fig2yes} but for
        run 2GH where the resolution is higher with
         88 instead of 56 cells between the two
         stars. The secondary wind is injected
         at a distance of 4 cells, but now it is
         closer to the secondary  as compared with run 2G.
         Consequently, the dense primary wind reaches somewhat closer to the secondary star. }
   \label{fig:fig2yesH}
     \end{figure}

We start by presenting some cases with an orbital separation of $r=2 \AU$.
In Fig. \ref{fig:fig2yes} we present run 2G at a particular time.
The entire wind collision region is bent toward the gravitating secondary star
(the star on the right), and a dense blob is clearly seen at a distance of $0.1 r$ from
the secondary.
This is taken as accretion, as it must be accreted under realistic conditions
(rather than an imposed secondary wind in the inner 4 cells).
To demonstrate the wiggling behavior of the winds collision process, in Fig. \ref{fig:fig2yest}
we show the same run but at nearly one month after the time of Fig. \ref{fig:fig2yes}.
This shows the wiggling of the winds collision region, and our imposed outflowing condition
near the secondary star.
If instead of imposing outflowing wind near the secondary, we would let the
primary wind to reach the secondary, we would had a continuous accretion.
Our limited resolution does not allow us to do so in the present study. In a future paper we will
concentrate on the accretion process itself.
In Fig. \ref{fig:fig2yesH} we present a run with gravity and the same parameters,
but with a higher resolution. The number of
cells between the two stars is 88 instead of 56,
but the injection of winds is still at a distance of 4 cells from each star.
As can be seen, the dense primary wind reaches somewhat closer to the secondary because the
injection region is smaller.
% FFFFFFFFFFFFFFFFFFFFFFFFFFFFFFFFFFFFFFFFFFFFFFFFFFF
% FFFFFFFFFFFFFFFFFFFFFFFFFFFFFFFFFFFFFFFFFFFFFFFFFFF
% FFFFFFFFFFFFFFFFFFFFFFFFFFFFFFFFFFFFFFFFFFFFFFFFFFF
% FFFFFFFFFFFFFFFFFFFFFFFFFFFFFFFFFFFFFFFFFFFFFFFFFFF
% FFFFFFFFFFFFFFFFFFFFFFFFFFFFFFFFFFFFFFFFFFFFFFFFFFF
% FFFFFFFFFFFFFFFFFFFFFFFFFFFFFFFFFFFFFFFFFFFFFFFFFFF

\begin{figure}
    \includegraphics[scale=0.6]{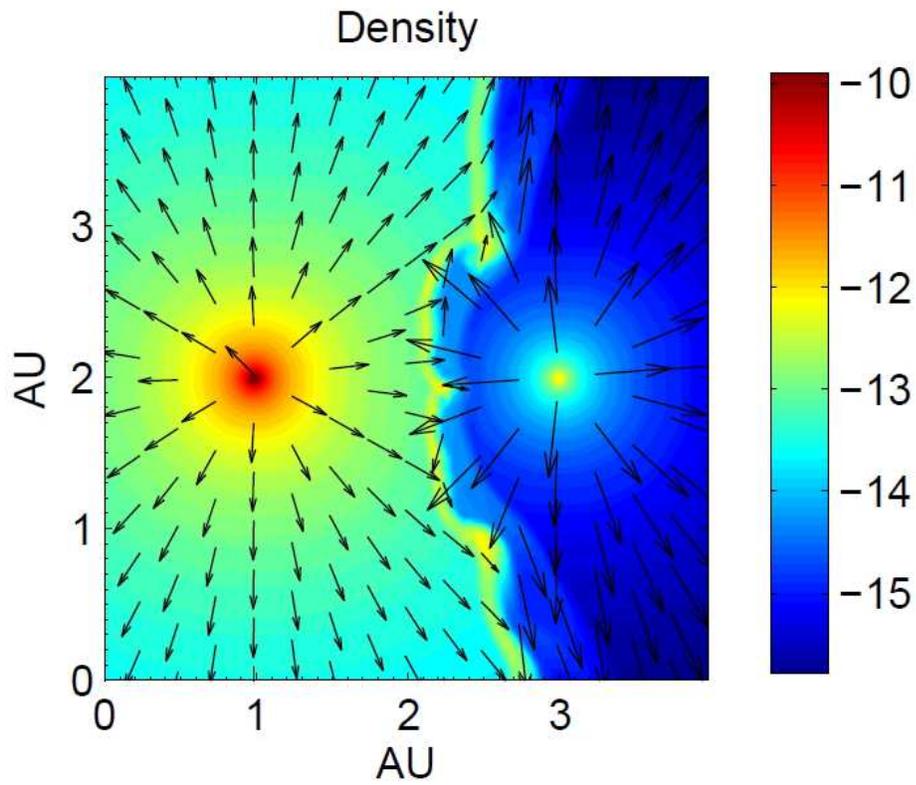}
        \caption{The run with $r=2 \AU$ but no gravity is included.
        Density and arrow scales are as in Fig. \ref{fig:fig2yes}.
        The differences from Fig. \ref{fig:fig2yes} are clear:
        the winds stagnation point is far from the secondary,
        and no accretion is taking place. }
   \label{fig:fig2no}
     \end{figure}

When the gravity of the secondary is not turn on (no gravity), no accretion takes place.
This is clearly seen from run 2N presented in Fig. \ref{fig:fig2no}.
The stagnation point (the point along the wind collision region where the
velocity is zero) is far from the secondary at all times,
and a general bow shock is seen.
No dense blobs from the shocked primary wind ever reach the
injection zone of the secondary wind.
The differences between the case with (Fig. \ref{fig:fig2yes}),
and without (Fig. \ref{fig:fig2no}), gravity are not only in the presence or not
of accretion, but the entire winds collision region is different.
This implies that simulations of the wind collision in $\eta$ Car that do not
include the secondary gravity are not adequate near periastron passage.
% FFFFFFFFFFFFFFFFFFFFFFFFFFFFFFFFFFFFFFFFFFFFFFFFFFF
% FFFFFFFFFFFFFFFFFFFFFFFFFFFFFFFFFFFFFFFFFFFFFFFFFFF

\begin{figure}
    \includegraphics[scale=0.6]{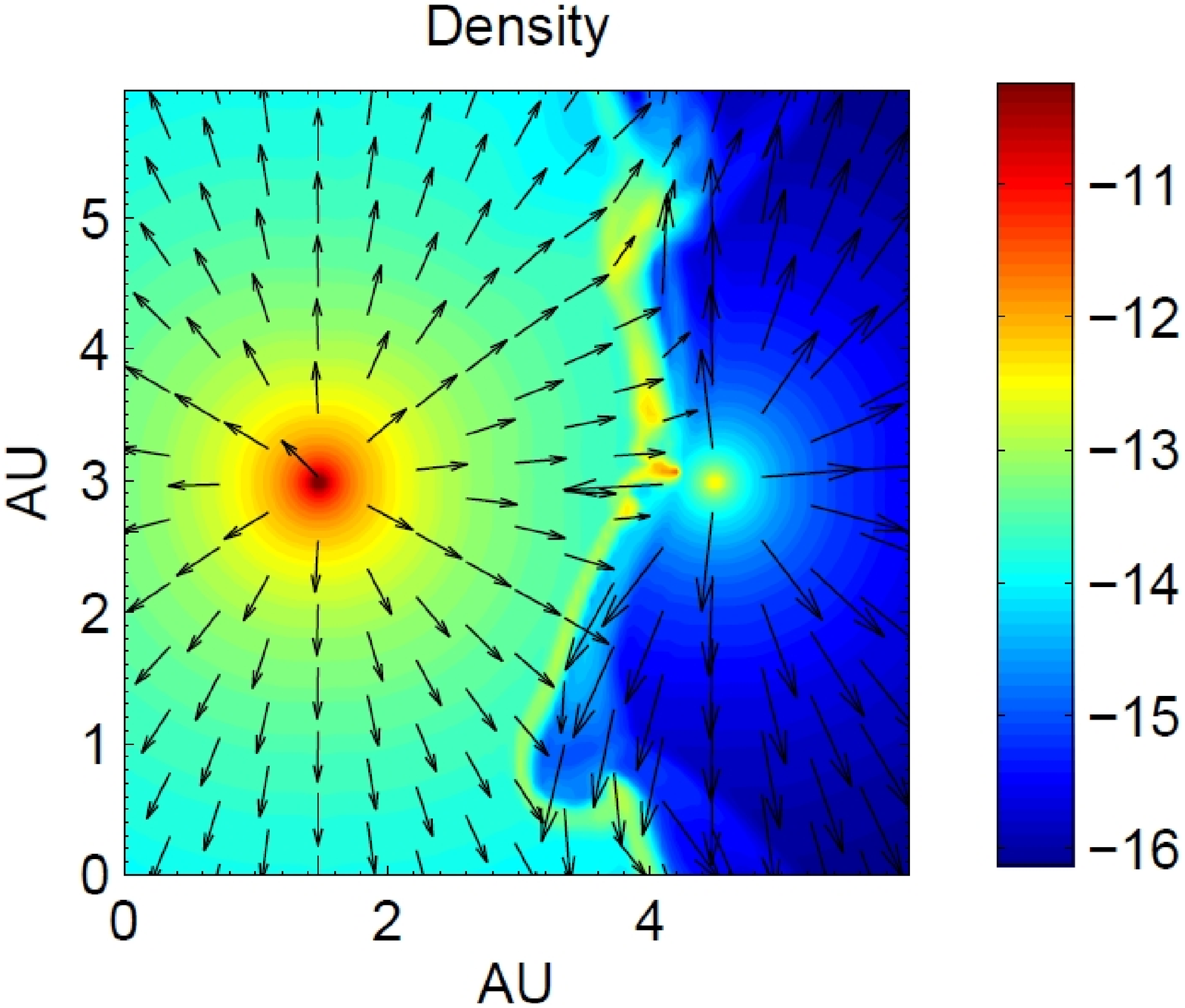}
        \caption{Like Fig. \ref{fig:fig2yes} but for
        an orbital separation of $r=3 \AU $. Accretion takes place.}
   \label{fig:fig3yes}
     \end{figure}

\begin{figure}
    \includegraphics[scale=0.6]{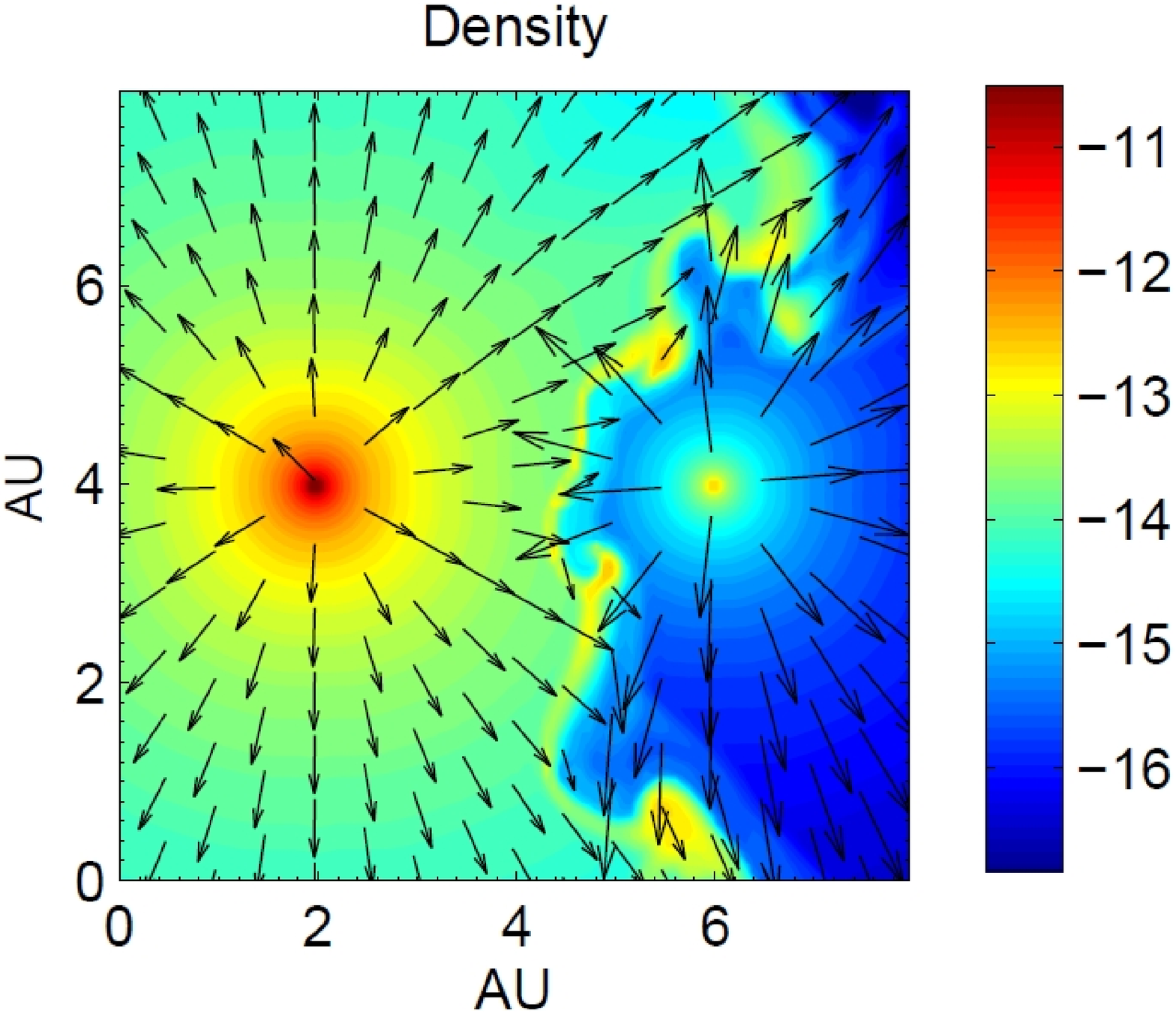}
        \caption{Like Fig. \ref{fig:fig2yes} but for
        an orbital separation of $r=4 \AU $. Accretion does not occur.}
   \label{fig:fig4yes}
     \end{figure}

To examine the approximate orbital separation where accretion
starts we present the cases for $r=3 \AU$ (run 3G) and $r=4 \AU$
(run 4G), in Figs. \ref{fig:fig3yes} and \ref{fig:fig4yes},
respectively. For an orbital separation of $r=3 \AU$ (about 17
days before periastron passage) accretions occurs, while when the
orbital separation is $r=4\AU$ (about 27 days before periastron
passage), there is no accretion onto the secondary star. As we do
not include the orbital motion and the acceleration zone of the
primary wind, there is no point it determining the exact orbital
separation where accretion starts under our assumptions. We can
say it starts at about an orbital separation of $\sim 3-4 \AU$, or
about three weeks before periastron passage. This is compatible
with the time that the X-ray emission starts to decrease (Corcoran 2005).
According to the accretion model, the accreted mass
suppresses the secondary wind (this is not shown here, and will be studied
in a future paper). After several more days, according to the model, the
suppression becomes significant, until the accretion process
(almost) completely shuts the secondary wind off.
% FFFFFFFFFFFFFFFFFFFFFFFFFFFFFFFFFFFFFFFFFFFFFFFFFFF
% FFFFFFFFFFFFFFFFFFFFFFFFFFFFFFFFFFFFFFFFFFFFFFFFFFF
% FFFFFFFFFFFFFFFFFFFFFFFFFFFFFFFFFFFFFFFFFFFFFFFFFFF
% FFFFFFFFFFFFFFFFFFFFFFFFFFFFFFFFFFFFFFFFFFFFFFFFFFF

\begin{figure}
   \centering
    \includegraphics[scale=0.6]{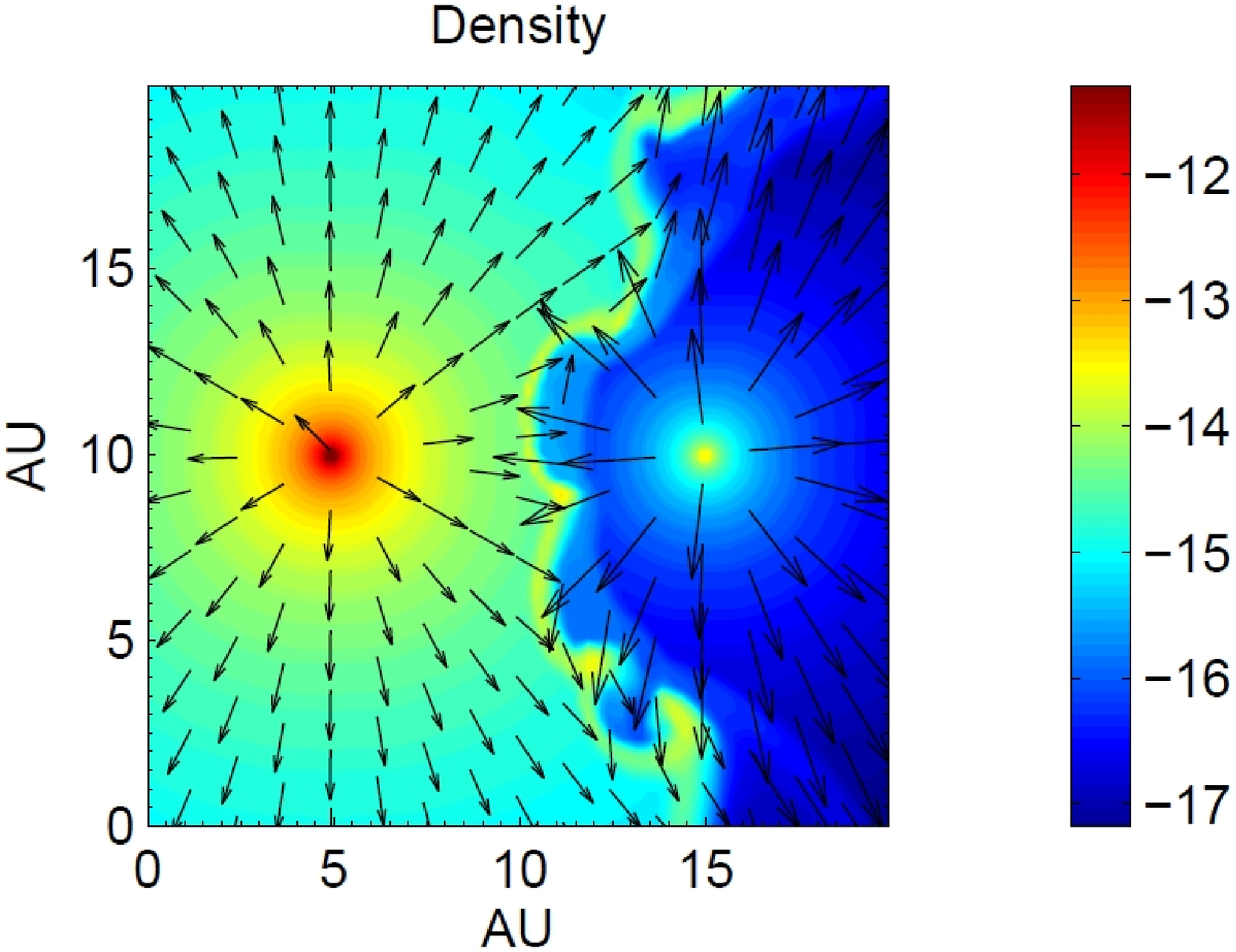}
    \includegraphics[scale=0.6]{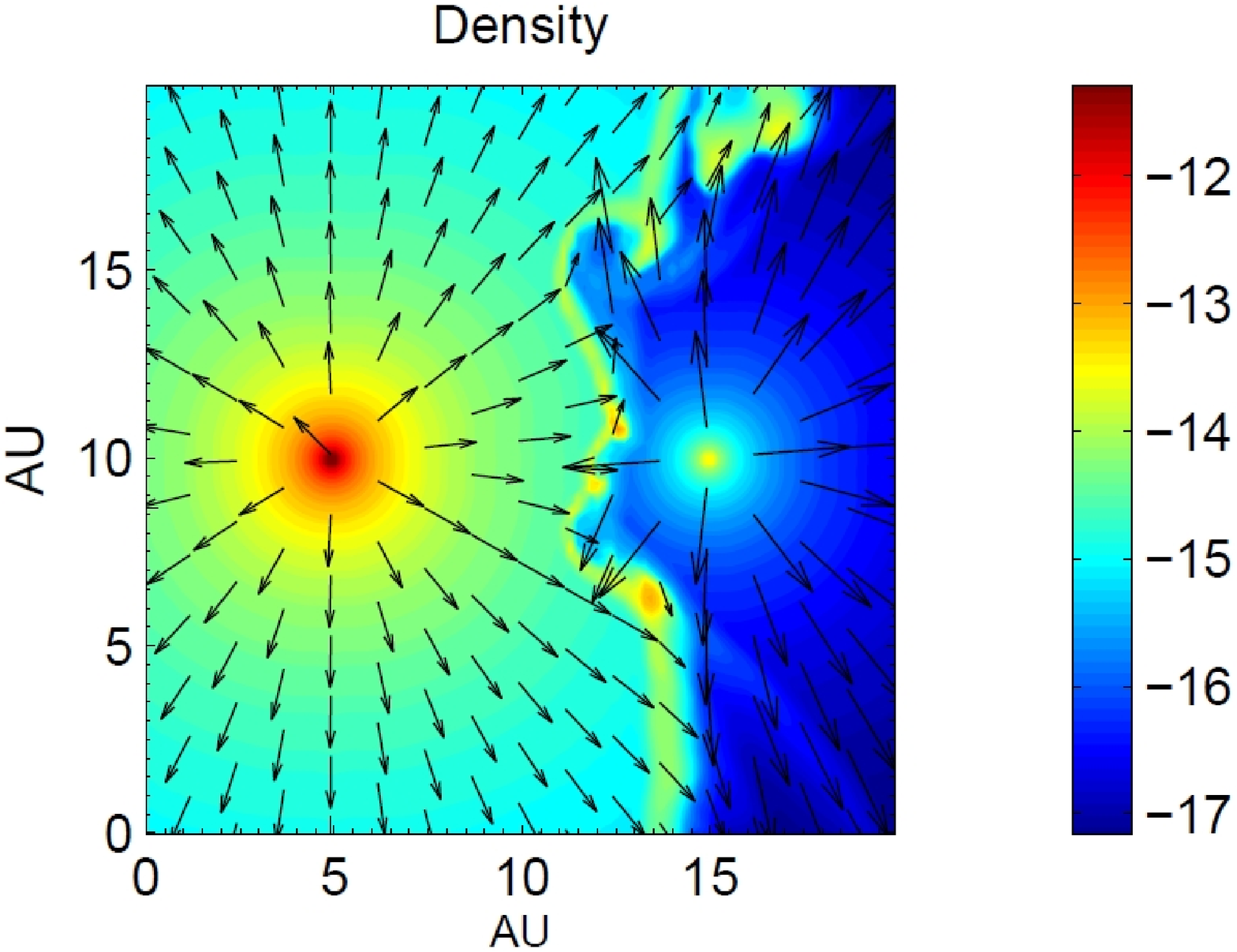}
        \caption{Winds collision with, upper panel, and without, lower panel, gravity and an
        orbital separation of $r=10 \AU $. Accretion does not occur.
        Comparison of the two panels shows that gravity plays a small role at
        larger orbital separations where $D_s \simeq 0.3r \gg R_{\rm acc}$. }
   \label{fig:fig10yesno}
     \end{figure}

In Fig. \ref {fig:fig10yesno} we also show two runs with (upper panel) and without (lower panel)
gravity, but at a large orbital separation.
It is evident that at large orbital separations gravity plays a small role.
In the case without gravity the stagnation point is even closer to the secondary.
The reason is that the interaction region wiggles, and
we show the temporary flow structure at a time
when in the run without gravity the interaction region is closer to the secondary.
The two length scales to compare are the accretion
radius $R_{\rm acc}= 2GM_2/v_1^2$, and the
distance from the secondary to the stagnation
point when gravity is not included $D_s \simeq 0.3 r$.
We find.
\begin{equation}
\frac{D_s}{R_{\rm acc}} \simeq 3
\left( \frac{D_s}{0.3 r} \right)
\left( \frac{r}{2 \AU} \right)
\left( \frac{M_2}{30 M_\odot} \right)^{-1}
\left( \frac{v_1}{500 \km \s^{-1}} \right)^2.
\label{eq:raccr}
\end{equation}
While at $r=10$ the accretion radius is
small, $R_{\rm acc}/D_s <0.1$, at $r \simeq 3 \AU$,
gravity is not negligible any more.
Including the orbital motion and acceleration zone of the primary wind will make accretion more
significant (Soker 2005; Kashi \& Soker 2009c).
% FFFFFFFFFFFFFFFFFFFFFFFFFFFFFFFFFFFFFFFFFFFFFFFFFFF
% FFFFFFFFFFFFFFFFFFFFFFFFFFFFFFFFFFFFFFFFFFFFFFFFFFF

\begin{figure}
   \includegraphics[scale=0.6]{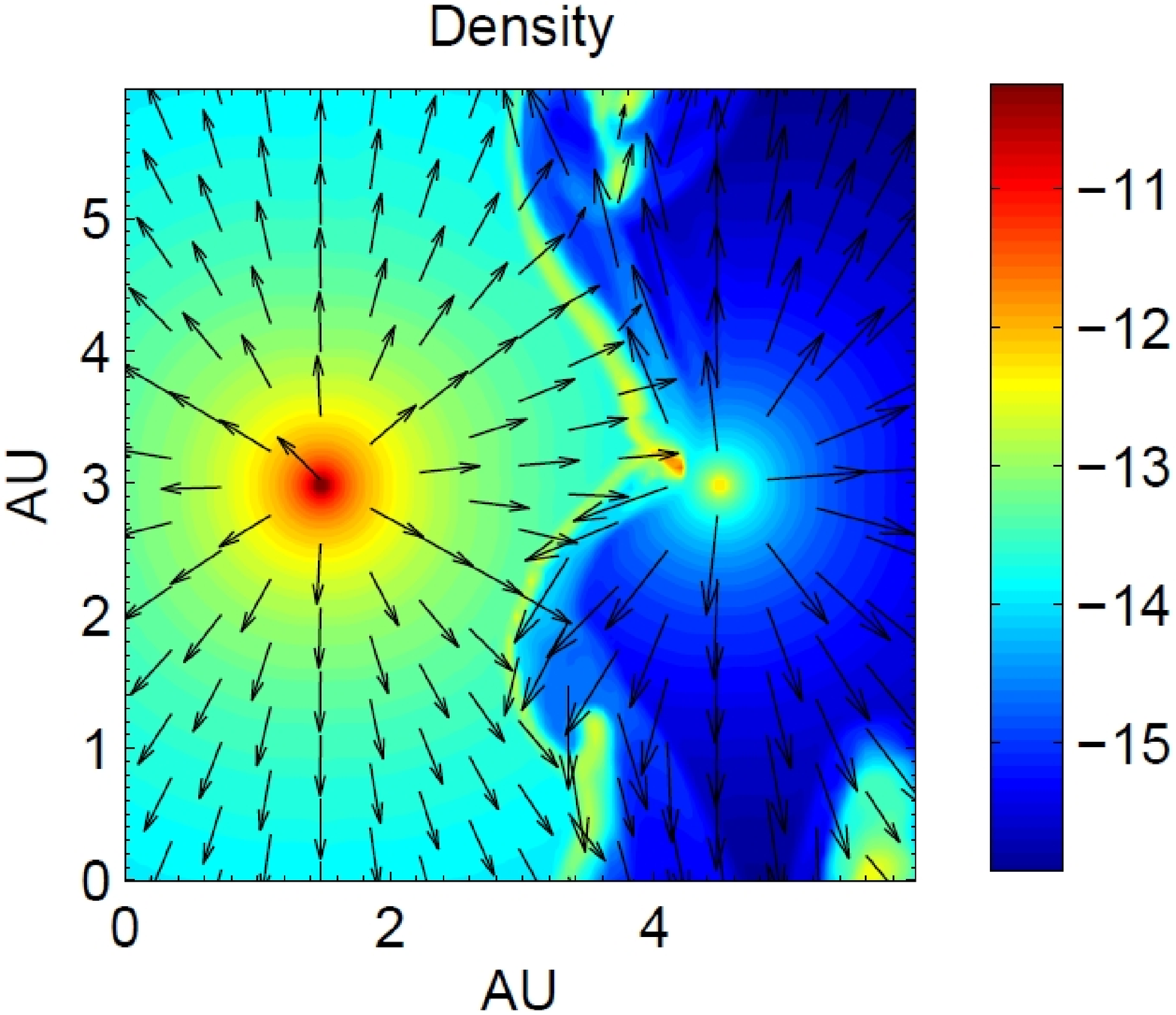}
        \caption{Run 3GR where $r=3 \AU$ and gravity is included,
        but the mass loss rate is $1.6$ time the mass
        loss rate in run 3G shown in Fig. \ref{fig:fig3yes}.
        The purpose is to mimic the secondary radiation pressure.
        Accretion is clearly seen.
        Density and arrow scales are as in Fig. \ref{fig:fig2yes}. }
   \label{fig:fig2yes16}
     \end{figure}

Taking the luminosity of the secondary as $L_2 = 9 \times 10^5 L_\odot$
(Verner et al. 2005), the magnitude of the radiation momentum rate, $L_2/c$,
is $(L_2/c)/(\dot M_2 v_2)= 0.6$ times
that in the secondary wind.
To mimic the radiation pressure, in run 2GR we take the secondary mass loss rate
to be 1.6 higher, $\dot M_2=1.6 \times 10^{-5} M_\odot \yr^{-1}$.
This is actually a `pessimistic' case, because the primary radiation pressure will
act in the opposite direction, and the dense primary wind does not absorb all
the secondary radiation. However, we want to test this case as well.
In Fig. \ref{fig:fig2yes16} we show the results of run 3GR. Clearly accretion is taking place.
% FFFFFFFFFFFFFFFFFFFFFFFFFFFFFFFFFFFFFFFFFFFFFFFFFFF
% FFFFFFFFFFFFFFFFFFFFFFFFFFFFFFFFFFFFFFFFFFFFFFFFFFF

We also present the temperature maps for three simulated cases.
The two wind are adiabatically cooling as they leave their respective stars.
They can be reheated if they are shocked.
The post shock primary wind is very dense, and its radiative cooling time is very short
down to our radiative cooling floor of $T=2 \times 10^4 \K$.
Adiabatic cooling is possible there after, but it is not significant.
As noticed by all papers that have studied the winds collision in $\eta$ Car
(e.g., Corcoran et al. 2001; Pittard \& Corcoran 2002; Akashi et al. 2006),
the post shock secondary wind has a long radiative cooling time.
The high-temperature regions of the post-shock secondary wind are clearly
seen in the three panels.
In the upper panel,  of the flow with $r=2 \AU$ where accretion occurs, the bending
of the winds collision region toward the secondary is clearly seen.
In the two other panels where accretion does not occur, the general bow
structure is seen.
% FFFFFFFFFFFFFFFFFFFFFFFFFFFFFFFFFFFFFFFFFFFFFFFFFFF
 \begin{figure}
 \centering
   \includegraphics[scale=0.8]{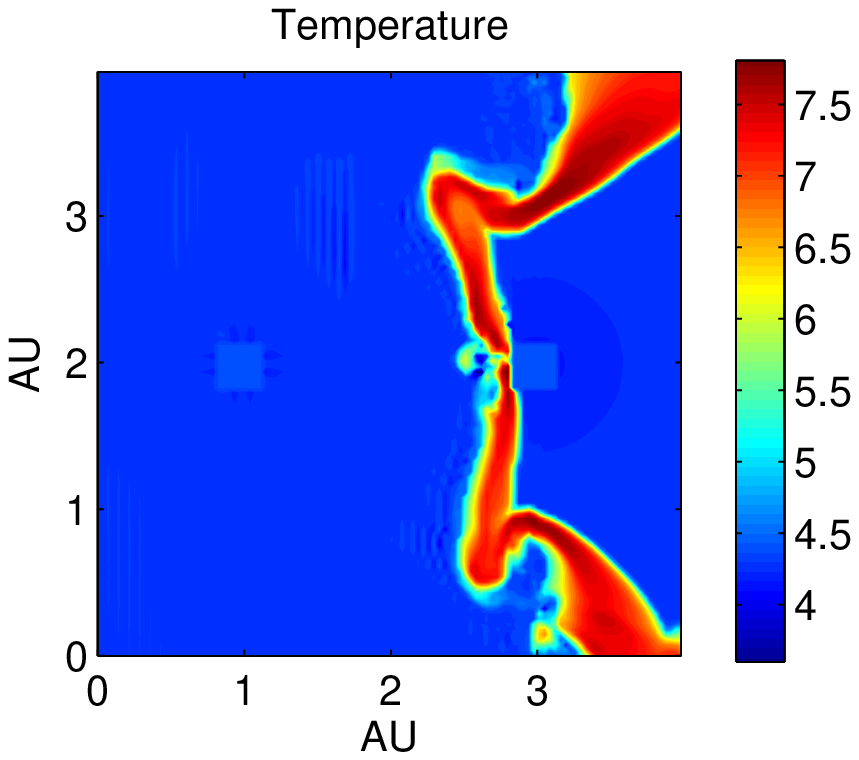}
   \includegraphics[scale=0.8]{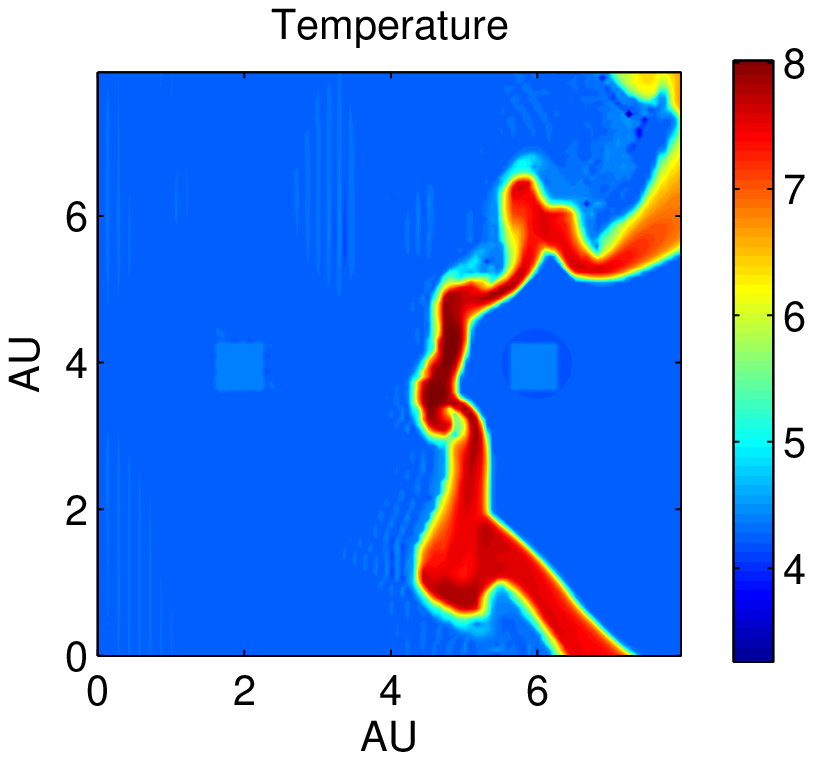}
   \includegraphics[scale=0.8]{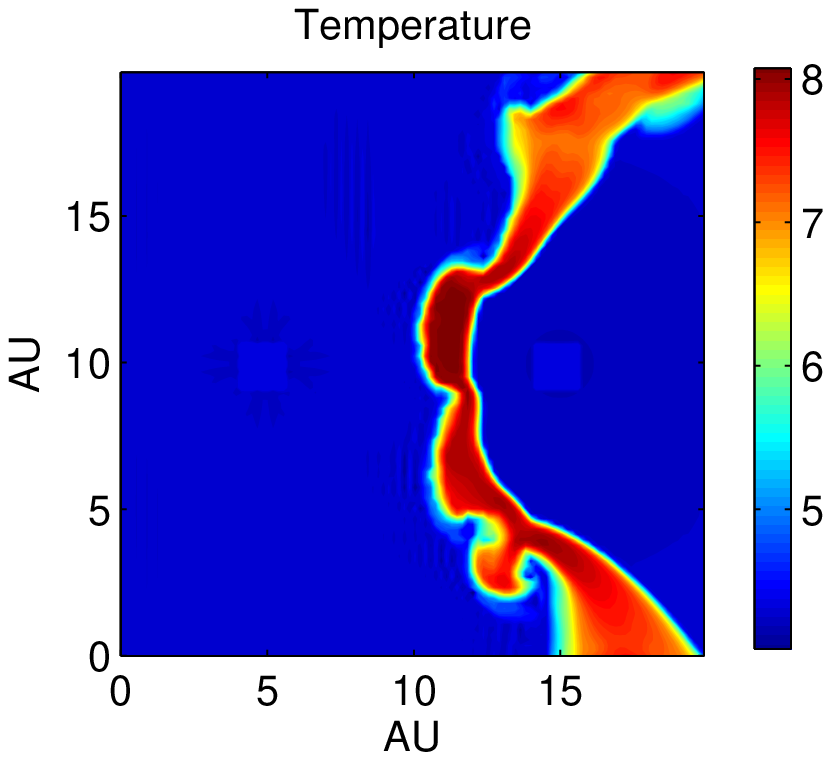}
        \caption{The temperature map of the runs: 2G, 4G, and 10G.
        The bar on the right gives the
        temperature color-code in units of $\log T(\K)$.
        Only in the flow shown in the upper panel accretion occurs.
        The secondary star is located at $(x,y)_2=(3,2)$, $(6,4)$, and $(15,10)$,
        from upper to lower panel.  }
   \label{fig:temps_map}
     \end{figure}
% FFFFFFFFFFFFFFFFFFFFFFFFFFFFFFFFFFFFFFFFFFFFFFFFFFF

We end by including the gravity of the primary star; runs 2GG and 3GG.
While the secondary wind velocity is much larger than the escape velocity from the secondary,
and more so from the injection radius that is larger than the secondary radius
($v_{w2} =3000 \km \s^{-1}$ compared with $v_{\rm esc}\simeq 750 \km \s^{-1}$),
the opposite is true for the primary star.
An escape velocity equals to the primary terminal wind velocity of $v_{w1}=500 \km \s^{-1}$
is reached at $r_1=0.85 \AU$.
The wind is continued to be accelerated by the radiation pressure of the primary
beyond this radius.
As we don't have the ability to include the radiation pressure, we set the
primary gravity to be zero inside a sphere of $r_{\rm G}=1 \AU$ and $r_{\rm G}=1.5 \AU$,
for the runs with an orbital separation of $r=2 \AU$ and $r=3 \AU$, respectively.
As the colliding region is outside the sphere, this does not influence the role
of the primary gravity in the relevant regions.

\begin{figure}
 \centering
  \includegraphics[scale=0.8]{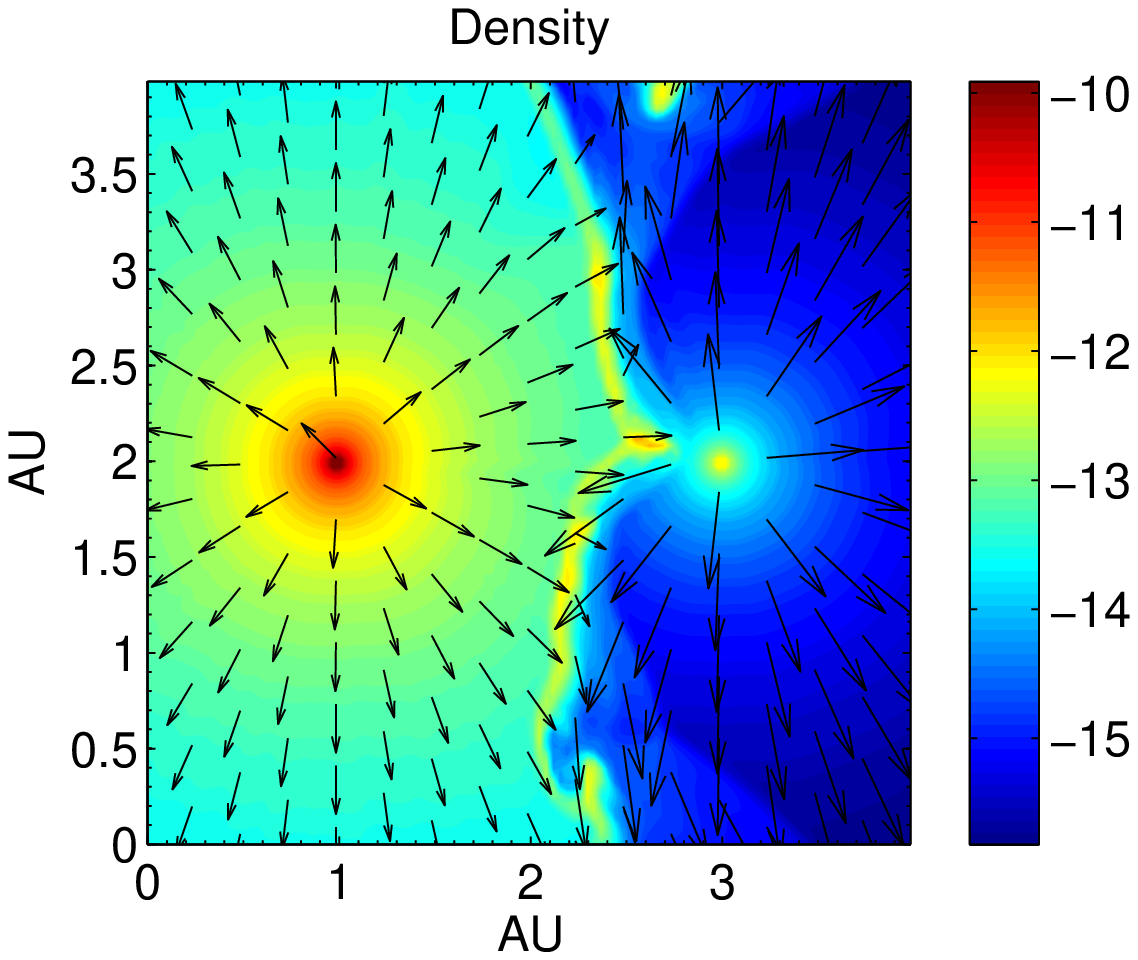}
  \includegraphics[scale=0.6]{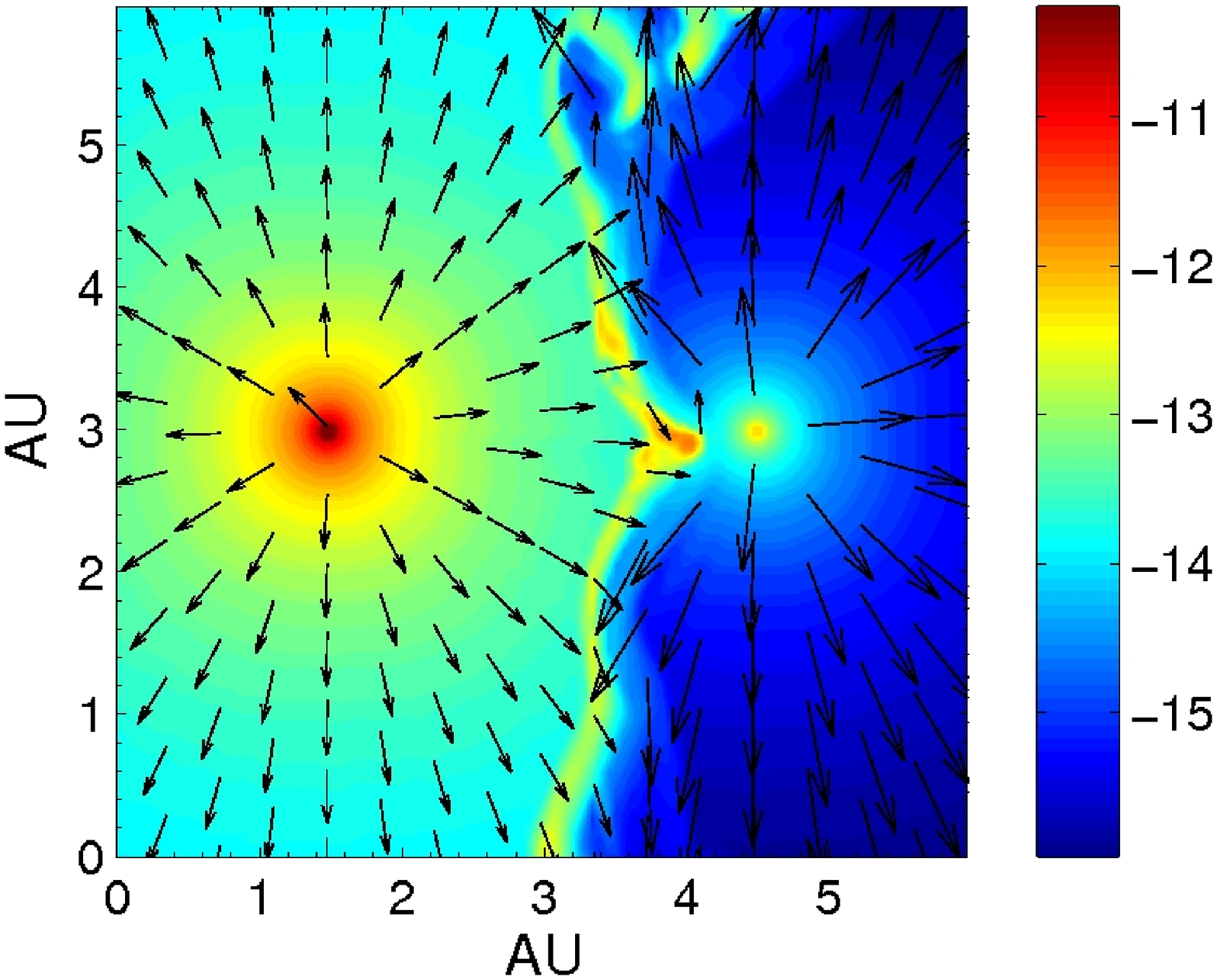}
        \caption{Like Figure \ref{fig:fig2yes}, but when the gravity of the primary star is included
        away from the primary (see text for detail).
        Upper and lower panels are for $r=2 \AU$ and $r=3 \AU$, respectively. }
   \label{fig:gravity1}
     \end{figure}

As can be seen in Fig. \ref{fig:gravity1} accretion does occur for both $r=2 \AU$ and $r=3 \AU$.
A dense blob near the stagnation point is seem to be located very close to the secondary star,
with a bridge of dense material extending to the secondary star. This is our criterion for accretion.
The structure of the colliding winds is a little different than the runs with no
primary gravity (Figs. \ref{fig:fig2yes} and \ref{fig:fig3yes}), in that the dense region opening is somewhat larger.
Namely, the primary gravity opens the colliding region.
% FFFFFFFFFFFFFFFFFFFFFFFFFFFFFFFFFFFFFFFFFFFFFFFFFFF
% FFFFFFFFFFFFFFFFFFFFFFFFFFFFFFFFFFFFFFFFFFFFFFFFFFF

%% ========================================
\section{SUMMARY}
\label{sec:summary}
% ========================================

 We conducted 3D numerical simulations of the winds collision process in the massive binary
 system $\eta$ Carinae.
Our numerical code includes radiative cooling of the two winds and the gravity of the secondary star.
In two runs we include the primary gravity. this does not prevent accretion.
However, with no radiation pressure the addition of the primary gravity is not self consistent.
Including radiation pressure will make accretion more likely even.

We reproduced features that have been observed in previous simulations
(Pittard et al. 1998; Pittard \& Corcoran 2002; Okazaki et al. 2008; Parkin et al. 2009),
such as the erratic motion of the winds collision region (wiggling), and a very hot region
of the post-shock secondary wind.
Our new numerical finding is that when the binary system reaches an orbital separation of $r \sim 3-4 \AU$,
about three weeks before periastron passage, accretion of dense primary wind gas onto the secondary star begins.
Because of the several simplifying assumptions made here, we cannot determine the exact separation
when accretion starts in reality.
The main assumptions are that we neglect the acceleration zone of the
primary star, we neglect the ram pressure of the primary wind due to the orbital motion,
and we conduct each simulation at a constant orbital separation.
In addition, we could not treat properly (because of limited resolution) the accretion processes
after the dense primary wind gas reaches the secondary star for the first time.

Although this is the first time accretion was found numerically for the case of $\eta$ Car, this is not
a surprising result.
Accretion is expected from simple analytical calculations (Soker 2005; Kashi \& Soker 2008b, 2009a).
We hope our results will encourage other researchers to include the secondary stellar gravity.
This will enable a better understanding of the processes that are behind the behavior of the system
near every periastron passage (the so called spectroscopic event).
The next periastron passage will take place in July 2014, leaving us plenty of time for a deep
theoretical study of the accretion process before the next event.

 We thank Amit Kashi for helpful comments.
 This research was supported by the Asher Fund for Space Research
 at the Technion, and the Israel Science foundation.

 \end{document}